\documentstyle[pra,aps,epsfig]{revtex}
\begin{document}
\title{Unzipping Dynamics of Long DNAs}
\author{Simona Cocco$^1$, R{\'e}mi Monasson$^{2,3}$ and John F. Marko$^4$}
\address{$^1$ Laboratoire de Dynamique des Fluides Complexes, 3 rue
de l'Universit\'e, 67000 Strasbourg, France.\\
$^2$ Laboratoire de Physique Th\'eorique de l'ENS, 24 rue Lhomond,
75005 Paris, France.\\
$^3$ Laboratoire de Physique Th\'eorique, 3 rue
de l'Universit\'e, 67000 Strasbourg, France.\\
$^4$ University of Illinois at Chicago, Department of Physics,
845 West Taylor Street, Chicago IL 60607-7059.}
\maketitle

\begin{abstract}
The two strands of the DNA double helix can be `unzipped' by application
of $\approx 15$ pN force.  
We analyze the dynamics of unzipping and rezipping, for the case where
the molecule ends are separated and re-approached at constant velocity.
For unzipping of 50 kilobase DNAs at
less than about 1000 bases per second, thermal equilibrium-based
theory applies. 
However, for higher unzipping velocities, rotational viscous 
drag creates a buildup of elastic torque to levels above $k_B T$ 
in the dsDNA region, causing the unzipping force
to be well above or well below the equilibrium unzipping force during
respectively unzipping and rezipping,
in accord with recent experimental results of Thomen {\em et al.}
[Phys. Rev. Lett. {\bf 88}, 248102 (2002)].  
Our analysis includes the effect of sequence on unzipping and rezipping, and
the transient delay in buildup of the unzipping force due to the approach
to the steady state.
\end{abstract}

\date{\today}

\section{Introduction}

Double stranded DNA (dsDNA, the DNA double helix) is the genetic
memory element of all cells.  Two copies of the genetic information
are encoded into the two complementary-sequence strands, which
are base-paired together through most of the cell cycle.
However, the two strands must be completely
separated during DNA replication, and partially separated during
DNA transcription.  In cells, the separation of DNA strands occurs
via forces applied by DNA-processing machinery.  Force-driven
dsDNA `unzipping' is therefore of direct biological relevance.

A few groups have carried out single-molecule studies of DNA unzipping
by force.  Essevaz-Roulet, Bockelmann and Heslot \cite{1,2} have studied
the $\approx 15$ pN forces encountered during unzipping of 50 kb $\lambda$ DNA.
Variations of the unzipping force with sequence were observed,
which are over the range of about 10 to 20 pN.  Other groups
have carried out similar experiments on unzipping of DNA \cite{3,4} and on RNA
helix-loop structures \cite{5}, observing similar unzipping forces.

A number of theoretical works \cite{6,7,8,9,9bis,10,11,12} 
have addressed the equilibrium statistical mechanics of dsDNA 
unzipping, with particular emphasis on the effects of sequence.  
Unzipping driven by DNA torque (`DNA unwinding') has drawn much less
attention in spite of elegant experiments \cite{13}
and corroborating theory \cite{14}.  
As a result, theoretical consideration of the
combined effects of force and torque on unzipping have only recently
been discussed \cite{11}.  Since DNA unzipping involves rotation of
the remaining double-helical DNA (Fig. 1), one expects that rotational drag
torque should produce a contribution to the force needed to unzip
DNA.  Experimentally, no dependence
of force on the rate of unzipping up to about 1000 base pairs per second 
(bp/s) has been observed.  However, in recent experiments of Heslot
an appreciable increase (up to $40\%$) in unzipping force was observed,
at unzipping rates in the range of 10 kbp/s\cite{heslot}.

This paper presents theoretical analysis of the velocity dependence of the
unzipping force, for large molecules where the kinetics is dominated by viscous
effects.  To do this we introduce a dynamic model of unzipping which combines
the polymer stretching dynamics of the extending single-stranded DNA (ssDNA)
regions, the rotation of the remaining double-stranded (dsDNA) region 
due to the unwinding generated by the unzipping \cite{15,16,17}, 
and the kinetics of the translation of the `fork' separating the 
ssDNA and dsDNA regions.
Although some work has been done on the Langevin dynamics of DNA unzipped
by constant tension \cite{8,9bis,11}, a number of open questions remain.
These include the unzipping force for constant end-to-end displacement
velocity, the effect of sequence, and the role of rotation of the dsDNA.

Below, kinetic equations for unzipping show that beyond a 
certain unzipping rate, the predictions of the theory of
equilibrium unzipping cease to apply.
For a 50 kb dsDNA this critical rate is about 1000 bp/s, similar
to the threshold seen experimentally\cite{heslot}.
In our theory there is a buildup of elastic torque in the dsDNA
due to the drag torque associated with the dsDNA rotation. 
The ssDNA stretching degrees of freedom and the fork
region itself remain near to equilibrium at experimentally accessible
unzipping rates.
We show that the relation between the unzipping force and the
elastic torque  buildup during opening
is simply determined by the
equilibrium  coexistence between closed and opened and stretched base pairs,
described by the binding-unbinding force-torque `phase diagram'.

We first describe the experimental situation in Section II, and then
in Section III we review the equilibrium theory of DNA unzipping, using a
mean-field approach.  We discuss the effects of force applied
to the ssDNA ends, and torque applied `upstream' to the dsDNA for 
homogeneous and heterogeneous sequences.
In Sec. IV we discuss the relaxation of ssDNA stretching and dsDNA twisting, 
and we then present a dynamic model for the propagation of the ssDNA-dsDNA
`fork' region for homogeneous and heterogeneous sequences.
We first present a simple theory where we assume
that the dsDNA twist is in a steady state.
The problem of rezipping of a dsDNA is also considered,
and it is shown that for rapid retraction, a left-handed
viscous torque delays recombination of the ssDNAs. 
Finally in Sec. V we analyze the DNA twist dynamics
in order to understand the delay in force buildup 
observed experimentally at high unzipping rates\cite{heslot}.

\section{Experimental setup}
\label{exp}

The experiment of Thomen, Bockelmann and Heslot\cite{heslot} unzips a
$\lambda$ dsDNA of 48502 base pairs (bp) in 10 mM phosphate buffer,
150 mM NaCl, Ph 7, at room temperature.  The $\lambda$ dsDNA is attached
to two dsDNA linker arms each of 7000 bp; one of these is anchored
to a movable glass slide, and the other is attached to a 
$\approx$-micron-diameter silica bead which is held in a laser trap (Fig.~1). 
The displacement of the glass slide at
a controlled velocity in the range 1 to 20 $\mu$m/s
forces the molecule to open. 
After the molecule is opened (typically after
$\approx 25\;\mu$m of displacement) the stage motion is inverted, 
allowing the molecule to rezip into dsDNA (reannealing).
The force transmitted to the ends of the linkers during the experiment
is measured using the position of the bead in the laser trap, which
is of stiffness $k_{opt} \approx 0.25$~ pN/nm.

The stiffnesses of the two dsDNA linkers
and the two ssDNAs depend on force, but are known from other
experiments on dsDNA and ssDNA.
At forces near 15 pN the total stiffnesses of the two dsDNAs 
(together 14000 base pairs) 
is $k_{ds}\approx  0.1$~pN/nm; the stiffness of the two ssDNAs
depends on the number of unzipped base pairs $n$, and is
$k_{ss}\approx 70/n$~pN/nm. 
Below, we will neglect the $\approx 0.1\;\mu$m shift of the bead
in the laser trap. 

\section{Static Unzipping}
\label{static}

The basic unzipping force follows from the simple argument that unzipping
will be thermodynamically favorable when the free energy required
for unzipping one base, $ 1\leq g\leq 4 k_B T$, is equal to the 
mechanical work done, $f \ell$, where $\ell \approx 1 $ nm is the 
projected length of ssDNA liberated during unzipping of one base pair.  
The resulting force $f \approx g/\ell$ is on the order of 10 pN  
as it has been observed experimentally. 
Below we describe static unzipping in more detail, focusing on
the force and number of opened base pairs at fixed displacement.
The equilibrium theory of unzipping describes experiments 
for velocities small enough to allow the system to stay in thermal 
equilibrium.

\subsection{Homogeneous sequence}
We start by considering an idealized homogeneous sequence, with a uniform   
$g_0 = 2.5 k_B T$, the averaged value on the $\lambda$ sequence.
The free energy cost $g$ per opened base pair
may be obtained for the $\lambda$ sequence, 
for the experimental conditions described above,
using the Mfold program \cite{Zuk00} with 
stacking and pairing free energies measured by Santa Lucia \cite{San98}.

To describe unzipping of the dsDNA we write down the work
that must be done by the force to separate the ends of the linkers
by a distance $2x$.  
This includes the work done by the force
to extend the two dsDNA linker arms by $2x_{ds}$,
the free energy cost of opening $n$  base
pairs of the $\lambda$ DNA, and the work done by the force to 
keep the ends of the ssDNA regions separated by $2(x-x_{ds})$,
and is therefore a function of the number $n$ of opened base pairs,
the extension $2 x_{ds}$ of the dsDNA linker arms, and the total
end-to-end distance $2x$.  We also include the work done 
by torque $\Gamma$ applied to the end of the dsDNA region;
this will be essential to considering the drag opposing rapid rotation
of the dsDNA region.
This free energy reads:
\begin{equation}
\label{enlib}
F_x(n,x_{ds})  = 2\;W_{ds}(x_{ds}) +2 W_{ss}(x-x_{ds},n) 
+n\, [g_0  + \Gamma \theta_0 ] 
\end{equation}
Here $W(x)=\int_0^x f(x') dx' $ is the work done by the stretching force $f(x)$
at fixed extension, for the dsDNA linker arms and the unzipped ssDNA, 
and $\theta_0 = 2\pi/10.5 = 0.60$
is the number of radians of rotation made during opening of each base pair;
this model is discussed in more detail in Ref. \cite{11}.

In the range of forces of 10-30 pN the dsDNA is extended enough that
the leading contribution from its high-force entropic elasticity, plus
linear stretching elasticity, give an accurate model:
\begin{equation}
\label{odjik}
x_{ds}(f)=
 L_{ds} \left [ 1-\frac 1 2 \left(\frac{k_B T}{f A}\right)^{\frac 1 2} 
+ \frac{f}{\gamma_{ds}}\right]
\end{equation}
We use persistence length $A=48$ nm and Young modulus 
$\gamma_{ds}=1000$ pN as determined by separate experiments\cite{Wang}.
The length of each 7 kbp linker is
$L_{ds}=2.38\; \mu$m.

The ssDNA is described by the freely-jointed-chain-like (FJCL) model\cite{18},
which gives the extension of $n$ of the unzipped bases as:
\begin{equation}
x_{ss} = n\; l_{ss}(f)
\end{equation}
where
\begin{equation}
\label{fjcl}
l_{ss} (f)= d\,\left[\coth\left(\frac{f\,b}{k_B T}\right)-
\frac{k_B T}{f\, b}\right]\,\left[1+ \frac{f}{\gamma_{ss}}\right]
\end{equation}
The monomer length $d=0.56$ nm, the segment length $b=1.4$ nm, and the 
stretching elastic constant $\gamma_{ss}=800$ pN  
are fitted from experimental data \cite{18}.
Eqs. (\ref{odjik}) and (\ref{fjcl}) can be used to obtain free energies via
integration by parts:
\begin{equation}
W(x)= x\,f(x) -\int_0^{f(x)} x(f')\; df'.
\end{equation}

Given the $W$'s and the total half-extension $x$,
the minimum of the total free energy with respect to linker
arm extension $x_{ds}$, $\partial F_x(n,x_{ds})/\partial{x_{ds}} =0$,
equilibrates ssDNA and linker tensions.
Then, minimization of the free energy with respect to the number of 
opened base pairs $n$, $\partial F_x(n,x_{ds})/\partial n =0$, 
determines the equilibrium unzipping force $f_u$, via:
\begin{equation}
\label{fu}
2 w_{ss}(f_u)  - \Gamma \theta_0 =g_0
\end{equation}
where $w_{ss}(f)\equiv\int_0^{f} l_{ss}(f')\;df'$.

Eq. (\ref{fu}) is a first-order transition coexistence condition,
stating that work done by the force and torque in opening a base pair
equals the base pairing free energy.  Fig. 2 shows the solution of
(\ref{fu}) plotted in the torque-force plane.  Note that
overwinding torque in the dsDNA ($\Gamma > 0$) increases the unzipping force.
The size of dsDNA torque needed to appreciably shift up
$f_u$ is $g_0/\theta_0 \approx 4.2 k_B T$; a left-handed
(negative, corresponding to dsDNA unwinding) torque of this magnitude
makes unzipping occur for zero force, close to the unwinding torque
inferred from experiments \cite{13}.  Below we will calculate how the
overwinding transiently built up in the dsDNA during rapid unzipping,
resulting from rotational friction upstream of the unzipping `fork',
will boost the ssDNA tension.

The number of unzipped base pairs is simply calculated from the
condition that the total displacement $2x$ is equal to the extension
of the two linkers arms $2x_{ds}$ plus the extension of the two single
unzipped single strands $2 x_{ss}$:
\begin{equation}
\label{fnx}
x=n\,l_{ss}(f)+ x_{ds}(f)
\end{equation}
At the beginning of an unzipping experiment,
the two linker arms first stretch until the extension $x_{dsu}\equiv
x_{ds}(f_u)$ is reached. 
At this point the $\lambda$ DNA starts to unzip, and the force
stays pegged at $f_u$, with the number of opened base pairs
proportional to further displacement:
\begin{equation}
\label{neq}
n _u (x)= \frac{x -x_{ds}(f_u)}{l_{ss}(f_u)}
\end{equation}

The equilibrium force and number of unzipped base pairs 
for this homogeneous model, with $g_0=2.5 k_B T$ 
and at zero torque, are plotted in Fig.~3.
The critical unzipping force is $f_u \approx 16$ pN,
and the average projection of each unzipped base along the unzipping 
direction is $l_{ss}(f_u)= 0.44$ nm.  This is close to what 
is observed experimentally\cite{heslot} at small displacement velocities 
$v\leq  1\ \mu$m/s; as unzipping proceeds the ssDNA is stretched to about
50\% of its total contour length of $\approx 1$ nm/base.

Given accurate knowledge of the elasticity of ssDNA,
unzipping experiments can determine the pairing free energy 
at room temperature.  For the homogeneous model, we find $g_0 = 2.5 k_B T$.
Prior to these experiments, this free energy difference was
indirectly inferred from model free energies obtained from
study of DNA melting at temperatures of 20 to 40 degrees above room 
temperature\cite{20}.  
A sequence-averaged point of view as presented above 
can give a rough account of thermodynamics of unzipping of large molecules.
However, there is appreciable sequence-dependence
of the base-pairing free energy.
Rief {\em et al} have found that pure AT (the most loosely
bound base pairs) sequences unzip at
about 9 pN, while pure GC sequences (the most tightly bound base pairs)
open at about 20 pN \cite{3}.
This range of force corresponds to base-pairing free energies of 
0.8 to 3.8 $k_B T$ per base pair.  
Therefore, a more detailed analysis of unzipping, and especially
describing unzipping of short inhomogeneous sequences, requires models
which take into account sequence-dependence and the
cooperativity of strand separation\cite{20}.

\subsection{Heterogeneous sequence}
\label{heterostatic}
Sequence effects can be added by making $g_0$ a function of $n$.
The equilibrium opening of $\lambda$-DNA at zero torque has been 
theoretically analyzed by Bockelmann, Essevaz-Roulet and 
Heslot\cite{2}, who have numerically calculated the 
thermal average of the force and the number of opened base pairs. 
They included thermal fluctuations of the ssDNA and dsDNA regions using
a free energy of the form (\ref{enlib}) plus trap/cantilever elastic energy.
Here we show how to obtain essentially the same results, using a
preaveraging of $g_0(n)$.

The approach of the previous section computes the free energy of
a given number $n$ of opened base pairs, at fixed displacement $x$,
using (\ref{enlib}) and (\ref{fnx}).
We estimate the fluctuations of the ssDNA, the dsDNA and the 
laser trap/cantilever, at $f_u \approx 15$ pN 
using their combined stiffness\cite{Boc02}
$k_{tot} = [1/k_{ss}+1/k_{ds} +1/k_{opt}]^{-1}$.
This net stiffness decreases with the opening because  
of the inverse proportionality of $k_{ss}$ to the number of opened base pairs.
During opening of the first 5 kbp of $\lambda$, the extension
fluctuations in the length are $\approx 20$ nm,
corresponding to 20 base pairs.

To account for these fluctuations 
we have Gaussian-preaveraged the denaturation free energy $g(n)$
using a standard deviation of 10 base pairs.
To each  configuration of $n$ opened base pairs is associated 
a Boltzmann factor using
free energy~(\ref{enlib}) where the force is determined by the
condition~(\ref{fnx}). 
The thermal averages of the number of unzipped base pairs and 
of the force, as a function of 
the displacement  $2x$, are plotted in Fig.~4, for displacement after
opening up to $5\ \mu$m ($\approx$ 5 kbp).  

The result of this calculation is in good agreement with experiment
and the computation of Bockelmann {\em et al.}\cite{Boc02}. 
The average slope of the opening curve
gives the average extension of each unzipped segment along the
unzipping direction $2 l_{ss} = 0.95$ nm 
(one base pairs opens for each $0.95$ nm of displacement).
As in the experiment,
the sequence generate a stick-slip motion; the opening fork stalls at 
G-C-rich parts of the sequence, 
giving a sawtooth pattern in the force signal and a step pattern 
in the number of opened base pairs\cite{Boc02,9bis}.

\section{Unzipping Dynamics}

To describe the motion of the unzipping fork, we must consider
four physical effects.  First and second, we must consider the elongating
dynamics of the dsDNA linker arms, and the unzipped ssDNA. 
Third, we must worry about the propagation of twist down the dsDNA; 
each 10.5 bases unzipped forces one more full right-handed twist 
into the unzipped dsDNA region.
Finally, we must close the dynamical equations with a model for
the translation of the ssDNA-dsDNA fork in $n$.
For each of these processes we consider relaxational dynamics 
of the form:
\begin{equation}
\label{din}
\zeta \frac{\partial z}{\partial t}= - \frac{\partial F}{\partial z}
\end{equation}
Here $z$ can be the position of a monomer of ssDNA or of dsDNA,
the twist orientation $\theta$ of a monomer of the dsDNA being opened, or 
the number of unzipped base pairs $n$.  In each case $F$ is the 
free energy, while $\zeta$ is the relevant friction constant.
For translational motion of ssDNA or dsDNA monomers,
$\zeta \equiv 6\pi \eta a$; we take $a= 1$ nm.
For twist relaxation, $\zeta \equiv 4\pi \eta r^2$ where
$r$ is the dsDNA hydrodynamic radius, and where
$\eta= 0.001$ Pa$\cdot$s, the suitable value for aqueous buffer.
We use $r=2$ nm, twice the `bare' chemical radius of the
double helix; this generates force curves in accord with 
the experiment\cite{heslot}. Note that $r$ is the only adjustable parameter
of our theory; it is essentially a friction constant.

\subsection{DNA stretch and twist relaxation times}

Since ssDNA and dsDNA are appreciably stretched by the $> 10$ pN forces 
applied during unzipping, their dynamics is reasonably described
by local hydrodynamic friction.  Logarithmic corrections due
to long-range hydrodynamic coupling can be added to this discussion
but without major effect. 
To estimate the order of magnitude of the dsDNA linkers
and the ssDNA to equilibrate, we expand the stretching free energies
around the force of $f=15$ pN (the nonlinear elasticity
of the previous section is used) to obtain:
\begin{equation}
\label{ela}
F = \frac{K}{2}\;  \sum_n\ \left[z(n,t)-z(n-1,t)-z_0\right]^2
\end{equation}
where $z(n,t)$ is the position along the
unzipping direction of either the $n$th monomer of the dsDNA, or of the
unzipped ssDNA.
In these two cases the monomer stiffness are either
$K_{ds}=1400$ pN/nm (dsDNA) or $K_{ss}=140$ pN/nm (ssDNA).
Note that these should note be confused with the {\it polymer} stiffnesses
$k_{ds}$ and $k_{ss}$ discussed previously.

The dsDNA twisting free energy is well described
by the free energy of an elastic rod\cite{21} in the range of 
torques relevant to unzipping experiments.
This free energy may be written as in (\ref{ela}) where the 
degree of freedom $z(n,t)$ is the twist angle
of base pair $n$. The twist stiffness is $K \equiv k_B T C/\Delta^2$
where the base rise  $\Delta = 0.34$ nm converts base index $n$ to dsDNA 
contour length (note $k_B T C = 80\pm 20$ nm is the usual
elastic-rod twist rigidity \cite{21})

The longest relaxation time of (\ref{din}) with an elastic free 
energy (\ref{ela}) is
\begin{equation}
\label{rela}
t=\frac{\zeta\; {\cal N}^2}{k\; \pi^2}
\end{equation}
where either ${\cal N}\equiv N_{ds}=14000$ 
is the number of base pairs in the two linkers DNA, 
the number of unzipped base pairs ${\cal N}\equiv 2n$,
or the number of still zipped base pairs ${\cal N}\equiv N-n$.
Substituting the relevant stiffness $k$ and drag $\zeta$ (\ref{rela}),
we obtain $t_{ds}=3\cdot 10^{-4}$ s for dsDNA stretch relaxation,
$t_{ss}=(2n)^2\;\times 1.4\cdot 10^{-11}$ s for ssDNA stretch relaxation,
$t_{tw}=(N-n)^2\times \;2\cdot 10^{-12}$ s for dsDNA twist relaxation times.

Equilibrium will be reached for stretching or twisting
if the relevant relaxation time is less than the unzipping time 
1 nm$\times t_u \approx n/v$. For the maximum velocities we are
considering, $2v=20\ \mu$m/s, $t_u= n \times 5 \cdot 10^{-5}$ s.
Since $t_{ss}/t_u \approx n/10^6$, ssDNA in unzipping experiments where
$2v \leq 20$ $\mu$m/s will be equilibrated until about $10^6$ bp are 
unzipped.  Therefore dsDNA and ssDNA stretching is
at equilibrium in experiments on $\lambda$-DNA ($n < 5 \times10^4$).

By contrast, twist
relaxation cannot reach equilibrium at the start of
unzipping; the relaxation time at $n=0$ is $t_0= 5 \cdot10 ^{-3}$ s.
Thus, we will now describe the fork dynamics,
treating the ssDNA and dsDNA stretching in equilibrium.
We first develop a theory of steady-state twisting; 
in Sec. \ref{twistprop} we analyze the approach to this steady state.
 
\subsection{Fork motion}
The fork position $n(t)$ will change as a result of the imbalance of
ssDNA tension, opening cost, and dsDNA torque.   Unzipping experiments occur at
less than 20 kb/s, slow enough that each base is opened on average
slower than single-base opening-closing times which are 
less than a microsecond.  A quasistatic model of fork motion is 
plausible, with fork velocity in proportion to the free energy change 
associated with unzipping of one base.  

We write the relaxation equation (\ref{din}) for the
fork motion in the continuum limit for the number of
opened base pairs:
\begin{equation}
\label{dinfor}
\tau_n {d n \over dt } = \frac 1{k_BT} \;\big( 2\, w_{ss} (f) -g_0(n) -\Gamma \,\theta_0\big)
\end{equation}
The single-base relaxation time $\tau_n$ should be on the order of the
diffusion time for the $\approx$ nm-long bases
$\tau_n \approx 6\pi\eta a^{3}/k_B T \approx 10^{-8}$ s.
The force $f(n,x)$ is again determined from (\ref{fnx}), and is
implicitly a function of both the displacement and the number of open bases.

Equation (\ref{dinfor})  indicates that if the ssDNA tension
is large, the fork moves to larger $n$.
The fork is static ($dn/dt=0$) for the equilibrium state (\ref{fu}).
Each unzipped base forces the upstream dsDNA to rotate through
$ \theta_0 = 0.63 $ rad.  If this fork rotation is sufficiently rapid,
the viscous rotational drag along the $N-n$  dsDNA
base pairs which remain to be unzipped, will generate elastic torque. 

In this subsection we assume that the DNA to be unzipped 
has reached a stationary state rotating
at a uniform angular velocity $\omega$.
Since each opened base forces a rotation of the dsDNA region by
an angle $\theta_0$, we have $\omega = \theta_0 dn/dt$.
The viscous torsional drag for the dsDNA, treated as a
cylinder of cross-sectional radius $r$ and length $\Delta (N-n) $, is:
\begin{equation}
\label{gamma}
\Gamma(n) = 4\pi\eta r^2 \Delta  \,( N - n )\; \omega
\end{equation}
Combining (\ref{gamma}) and (\ref{dinfor}) we obtain the equation
of motion for the fork position:
\begin{equation}
\label{dynan}
{d n \over d t} = \frac 1{k_BT}\; \frac{2\, w_{ss}(f) - g_0(n)}{\tau_n +(N-n) \tau_r}  
\end{equation}
where
\begin{equation}
\tau_r=\frac{4 \,\pi \eta r^2\; \Delta \;\theta_0 ^2}{k_B T} 
=2\cdot 10^{-9} \;{\rm s}
\end{equation}

The time $\tau_r$ is comparable to the value expected for $\tau_n$ (both are
viscous times at the nanometer scale), but since $\tau_r$
appears in (\ref{dynan}) magnified by a factor $\approx N$ relative to $\tau_n$,
the rotational dynamics will be rate-limiting in most experimental situations.
Also note that Eq. (\ref{dynan}), 
and $\tau_r$ are independent of the value of the
twist elastic constant $C$.
Numerical integration of (\ref{dynan}) gives force
and torque during unzipping.
The initial condition is that unzipping begins when the force 
in the linkers reaches $f_u$, i.e.  $n(t=x_{dsu}/v)=0$.

\subsection{Rezipping dynamics}

In the experiment of \cite{heslot}, following 
unzipping of $\approx 25000$ base pairs the molecule is allowed to
rezip (`reanneal' in the nomenclature of biochemistry), 
by reversing the direction of the pulling velocity.  
If velocity is made negative, (\ref{dynan}) describes this process.
During rezipping, the dsDNA rotates in a right-handed sense, 
generating a left-handed drag torque on the molecule.

As shown in the phase diagram
of Fig.~2 a negative torque promotes helix opening, decreasing the
unzipping force.  Therefore, the force during rezipping at high
velocities is lower than the equilibrium unzipping force $f_u$; 
the force drops progressively during rezipping because the
rotational drag (\ref{gamma}) increases with dsDNA length.
Force during retraction is calculated via
integration of (\ref{dynan}) (with $v\to -v$),
starting from initial condition $n=25000$.  

\subsection{Analytical estimate of unzipping force dependence on velocity}
We can estimate the increase of steady-state unzipping force with
velocity using (\ref{dynan}) for homogeneous
sequence $g_0=2.5 k_B T $.
Changing variables from $t$ to $x=vt$, 
(\ref{dynan}) can be rewritten as,
\begin{equation}
{d n \over d x} = \frac{1}{ (k_BT)\; v} \left[ 
\frac{2\, w_{ss}(f) - g_0}{\tau_n +(N-n) \tau_r}  \right]
\label{dynan2}
\end{equation}
where $f(x,n)$ is the equilibrium tension given by (\ref{fnx}).
In the limit of slow unzipping, the
square bracket of (\ref{dynan2}) goes to zero, giving the 
equilibrium unzipping force 
$f_u$ (\ref{fu}) and the relationship (\ref{neq}) between extension $x$ and 
number $n_u$ of unzipped base pairs, $n_u(x)=(x-x_{dsu})/l_{ss}(f_u)$.

We now suppose that the velocity is small (in a sense made more
precise in the following), and that the unzipping
force $f$ and number $n$ of unzipped base pairs can be expanded
to first order in $v$, i.e.
$f(x)=f_u +  f_1(x) \, v + O(v^2)$ and 
$n(x)=n_u(x) + n_1(x) \, v + O(v^2)$. 
Plugging this into (\ref{fu}) and (\ref{dynan2}) permits us to obtain two
coupled equations involving $f_1$ and $n_1$, with the results
\begin{eqnarray}
\label{force2}
f_1 (x) &=&  k_BT \left( \frac {\tau_n +[N-n_u(x)]  \tau_r}
{2\, [l_{ss}(f_u)]^2 } \right) \ ,\nonumber \\
n_1(x) &=& - \bigg[{x'_{ds}(f_u) + l_{ss}'(f_u) n_u(x)}\bigg]
\frac {f_1(x)} {l_{ss} (f_u)}
\quad ,
\end{eqnarray}
where $h'$ denotes the derivative of $h$ with respect to its argument.

Therefore, at low velocity, the unzipping force reads
\begin{equation}
f(x)= f_u \, \left( 1+ \frac {v}{v^*(x)} +O(v^2) \right)
 \ , \label{newforce}
\end{equation}
where 
\begin{equation} \label{newv*}
v^*(x) = \frac{2\,f_u \, l_{ss}(f_u)^2}{k_BT [ (N-n_u(x)) \tau _r + \tau
_n]} \quad .
\end{equation}
The function $v^*(x)$ is plotted in the inset of Fig.~5. It is a
rapidly increasing
function of displacement $x$, bounded from below by $v^*(x_{dsu}^+) \approx$
20 $\mu$m/s for $\lambda$-DNA ($N = 5\times 10^4$ bp). 
Therefore, one can expect to observe a large increase in the initial
unzipping force for velocities larger than a few $\mu$m/s.

Fig. 5 shows this force vs. velocity behavior 
which should be an upper bound to forces observed during the
unzipping of a $\lambda$-DNA (dashed line).  
The theory indicates a 10\% increase in unzipping
force as $v$ is increased to about 2 $\mu$m/s,
comparable to the initial rate of increase recently observed
\cite{heslot} (velocities
reported by Thomen {\em et al.} correspond to $2v$, see Fig. 1).

Eqs. (\ref{newforce}) and (\ref{newv*}) 
also show that as the molecule unzips and
$N-n$ goes down, the torsional drag on the dsDNA is reduced,
and the force needed to keep the fork moving goes down.  For
low velocity, this force drop will be a nearly linear function
of $N-n_u$.  This is the signature that the twist transport dominates
the fork retardation.
Finally, we note that (\ref{newforce})  can be
used to estimate how the force depends on velocity during
retraction by simply inverting the sign of the velocity.  
  
\subsection{Rezipping at zero tension}
For sufficiently fast retraction the reannealing of the double helix,
the rate of which will be limited by the rotational drag on the dsDNA region, 
will not be able to keep up with the retraction. 
As a result the force should essentially drop to zero.
This effect will be especially pronounced during the later stages of
rezipping, since the rotational drag experienced by the dsDNA region 
increases as rezipping proceeds.

The  velocity for `free' rezipping under zero tension can be
easily estimated, using
the equation of motion (\ref{dynan}) for the dsDNA-ssDNA fork:
\begin{equation}
{dn \over dt} = - \frac{1}{k_B T}\;  { g_0 \over \tau_n + (N-n)\,\tau_r }
\end{equation}
Taking initial condition $n(t=0)=n_0$, we obtain
\begin{equation} \label{re}
n(t) = N + {\tau_r \over \tau_n}- \left[ \left( {N+{\tau_n \over \tau_r}
-n_0}\right)^2 + 2\;\frac{ g_0}{k_B T}\;  {t \over \tau_r} \right]^{1/2}
\end{equation}
Recall that $\tau_n/\tau_r=5$ is not a large number. Therefore, for
long ($N > 10^3$) and initially totally open ($n_0=N$) molecules, 
the rezipping follows 
$N-n \approx (2\;g_0\;  t/(k_B T\;\tau_r))^{1/2}$, where the exponent is determined
by the linear dependence of the total rotational drag on the size
of the rezipped domain.

The total time needed to rezip for a `free' fork is therefore
$T \approx (2N-n_0)n_0\, \tau_r\;k_B T  /(2\; g_0) $, which for 
$n(0)=25000$ is $\approx 0.75$ s.
Therefore, for half-unzipped $\lambda$-DNA, retraction velocities of $2v 
\geq 50 $ $\mu$m/s are at essentially zero tension.
Equations (13) and (20) lead to the dsDNA torque for zero-tension rezipping:
\begin{equation}
\Gamma  = -{g_0\over\theta_0} \left[ 1+ {\tau_r \over \tau_n (N-n(t))}
\right] ^{-1} \quad ,
\end{equation}
with $n(t)$ given by (\ref{re}).
The maximum torque that appears at the fork
during zero-force rezipping is the critical torque for
the opening at zero force $g_0/\theta_0$, 
in accord with the phase diagram of Fig.~2.

\subsection{Results for homogeneous sequence}

We now present numerical results for 
integration of (\ref{dynan}) for the homogeneous case $g_0 = 2.5 k_B T$,
$\eta=10^{-3}$ Pa s, $r = 2$ nm, $\Delta=0.34$ nm, and $\theta_0=2\pi/10$.
The ssDNA elasticity FJCL parameters are
$b=1.4$ nm, $d=0.56$ nm and $\gamma_{ss}=800$ pN.
The dsDNA linker elasticity parameters are $L_{ds}=2.38 \mu$m,
$A=41$ nm and $\gamma_{ds}=1300$ pN. 
The fork and torsional relaxation times are
$\tau_n=10^{-8}$ s,$\tau_r=2\;10^{-9}$ s.    

Force as a function of displacement after opening is shown in Fig.~6 
for velocities $2v$ of 4, 8, 16 and 20 $\mu$m/s.
During unzipping one observes an initial force upswing
as the dsDNA linkers are first tensed, followed by a force
peak and a gradual force reduction due to the decrease of
torsional drag, as unzipping proceeds.
For $2v=20 \mu$m/s the peak force is 
$f_{max} \simeq 23$ pN, corresponding to a 7 pN unzipping force 
increase relative to the equilibrium value $f_u=$15.7 pN; this
is in good agreement with the increase of $\approx 10$ pN
observed in the experiment of \cite{heslot}.
On the other hand, the initial force increase observed experimentally 
is smoother than the theory of Fig. 6.  Below we will show how
sequence effects and twist relaxation dynamics reduce the initial
rate of force increase.

Fig.~6 also shows force during retraction.
For $2v=-4$ $\mu$m/s, there is already
a noticeable force hysteresis relative to the $2v=+4$ $\mu$m/s 
extension curve. For $2v\geq -8$ $\mu$m/s, the ssDNA force
approaches zero at the end of the retraction cycle.  Similar
`hysteresis loops' were observed by Thomen {\em et al.}\cite{heslot}.

Fig. 7 shows the DNA torque at the unzipping fork, during these
unzipping-rezipping cycles. 
The unzipping torque reaches a peak coincident with the force peaks of Fig. 6.
At $2v=20$ $\mu$m/s
the maximum torque during unzipping is 
$\Gamma_{max}=2.8 k_B \,T$, while the maximum
unwinding torque during rezipping is $\Gamma_{max}=-3.8 k_BT$,
slightly smaller in absolute value than the zero-tension limit
of $\Gamma=-g_0/\theta_0=-4.2 k_BT$.

The force as a function of torque during opening and closing run
at the maximal velocity of $2v=20$ $\mu$m/s
are included  in the phase diagram of Fig.~2. 
These curves follow the ssDNA-dsDNA equilibrium transition,
indicating that the opening and the closing of the single base pairs
takes place essentially at equilibrium.
The relation between torque and force at the moving fork
is, under likely experimental conditions, determined by the
base-pairing interactions in the same way as at equilibrium.
Points A,B,C,D in Fig.~6 and Fig.~7 are  mapped  to the coexistence
curve in Fig.~2;
for example, the force peak corresponds to 
point B in the phase diagram $f=23$ pN, $\Gamma= 2.8 k_BT$.  
Turning this around, this indicates that the
experimental force-displacement curve can be used to infer
the torque-displacement curve, using the equilibrium coexistence
line of Fig.~2.

The peak forces observed during unzipping are plotted in Fig. 5
(solid line), and match the approximation of the previous
section at low forces.  
Fig. 5 also shows the slightly lower force occurring
at the point where the molecule is 50\% unzipped (dot-dashed line).
This force is reduced simply because at the half-unzipped point,
there is less dsDNA remaining to provide torsional drag than
at the peak force point (see Fig. 6).

\subsection{Results for heterogeneous sequence}

Fig.~8 shows the results of numerical integration of
(\ref{dynan}) using the preaveraged
$\lambda$-DNA pairing free energy $g_0(n)$ as discussed in 
Sec.~\ref{heterostatic}.
The force as a function of the displacement for the velocities 
4,8,16,20 $\mu$m/s is in good agreement with experimental data,
and reflects the sequence; note, for example the progressive
increase in G-C percentage in the first 2000 base pairs,   
the decrease from the base pair 20000 to 24000 and the steeper
increase from base pair from 24000 to 24500, all these feature are
well reproduced in the experimental and theoretical curves. 
An interesting effect is that the fluctuations in the force 
due to the sequence are attenuated, especially during rezipping, 
at higher velocity.  

\section{Effect of twist relaxation}
\label{twistprop}
The initial increase of the force of Fig. 8 is still faster than
that observed experimentally.  We now examine the effect of
the initial twist relaxation dynamics, focusing on its influence
on the force signal at the beginning of unzipping.
The combined set of equations for the fork and twist comprises a
moving-boundary-condition problem that is difficult to solve
even by computation.
In this section, we construct an approximate solution for the
combined twist and opening dynamics, valid when the number of unzipped base
pairs remains small with respect to $N$. 

\subsection{Memory kernel for fork motion}

We rewrite (\ref{din}) and (\ref{ela}) as
\begin{equation}
\tau_{tw}\; \frac{\partial \theta}{\partial t} = 
\frac{\partial^2 \theta}{\partial m^2} \quad ,
\label{twist2}
\end{equation} 
where $m$ is the continuous base pair index, and where
$\tau_{tw} = 4\pi\eta r^2\Delta^2/C \simeq 1.4\, 10^{-11}$~s.
Prior to unzipping, the
dsDNA is relaxed,
\begin{equation} \label{initwi}
\theta (m,0) = 0 \qquad \qquad (0\le m\le N)\ .
\end{equation}
The boundary conditions expresses that the $m=N$ extremity of the molecule
is free (zero applied torque), while the location and unwinding
of the other extremity depend on the number $n(t)$ of unzipped
base pairs at time $t$,
\begin{eqnarray}
\label{extN}
\frac{\partial \theta}{\partial m} (N,t) &=& 0 \qquad , \\
\label{ext0}
\theta \big( n(t),t \big) &=& -\theta _0\, n(t) \quad .
\end{eqnarray}
We now introduce the Laplace transform of the twist,
$\bar \theta (m,p) = \int _0 ^\infty dt \, e^{-p t} \, \theta (m,t)$.
The solution of  (\ref{twist2}) with boundary condition (\ref{extN})
and initial condition (\ref{initwi}) reads
\begin{equation}
\bar \theta (m,p) = J(p) \; \cosh\big[ q (N-m) \big] \quad ,
\label{sol1}
\end{equation}
where $q^2=p\,\tau_{tw}$ and $J(p)$ has to be determined to fulfill
the remaining boundary condition (\ref{ext0}). Defining the Laplace
transform $\bar n (p)$ of $n(t)$, (\ref{ext0}) reads
\begin{equation}
-\theta _0 \; \bar n (p) = \int _0 ^\infty dt \, e^{-p t} \, \theta \big(
n(t),t \big) \simeq \bar \theta (0,p) 
\quad ,
\label{approxtwi}
\end{equation}
as long as $n(t)\ll N$ from (\ref{sol1}). Eliminating $J(p)$ from
(\ref{sol1}) and (\ref{approxtwi}), we obtain
\begin{equation}
\bar \theta (m,p) = -\theta _0 \; \bar n(p) \; \frac{ 
\cosh\big[ q (N-m) \big] }{\cosh\big[ q N \big] }\quad .
\label{sol2}
\end{equation}
From (\ref{sol2}), the derivative of the twist just upstream
of the fork, $\partial \theta/\partial m (n(t),t)$, determines
the fork torque. 
Inserting this into the Laplace transform of the equation
of motion (\ref{dinfor}) for the number of unzipped base pairs yields
\begin{equation}
\label{nofp}
\tau _n \; \bar n(p) = \frac 1{p+ a \, q \, \tanh(qN)}
\int _0 ^\infty dt \, e^{-pt} \bigg[ \frac{2 w_{ss} (f) - g_0(n(t))}
{k_BT} \bigg]
\ ,
\end{equation}
with $a=\theta _0^2 C/\Delta/\tau_n$.  
Inverse Laplace-transforming (\ref{nofp}),
we obtain a self-consistent integral equation for the
number of unzipped base pairs,
\begin{equation}
\label{noft}
n(t)= \int _0 ^t dt'\; G(t-t') \; \bigg[  \frac{2 w_{ss} (f)  - g_0
(n(t))} {k_BT} 
\bigg]
\quad , \label{integ}
\end{equation}
where the memory kernel $G$ is defined through
\begin{equation}
G(\tau ) = \frac 1{\tau _n} \int _{-\infty} ^\infty \frac{dp}{2\pi i} \;
\frac{e^{p \, \tau}}{p+ a \, q \, \tanh(qN)}
\quad . \label{glap}
\end{equation}
Let us stress that the force $f$ in (\ref{noft}) depends on $n$ and $t$
through (\ref{fnx}).
The poles of (\ref{glap}) are located on the real negative
semi-axis, at $p_\ell = -(y_\ell/N)^2/\tau_{tw}$;
$y_\ell$ is the root (unique) of $\tan y = - b \; y$
such that $|y _\ell -\ell \pi| < \pi/2$ ($\ell \ge 0)$, 
and $b\equiv a N  \tau_{tw}$. 
Calculation of the residues is straightforward, giving
\begin{equation}
G(\tau ) = \Theta (\tau) \left\{ \frac 1{1+b} + 2
\sum _{\ell =1}^\infty \frac{e^{-\tau/\tau_\ell}}{1+b+ y_\ell^2/b}
\right\} \quad , \label{glapp}
\end{equation}
where $\Theta (\tau)=1$ if $\tau>0$, 0 otherwise.
Finally,
\begin{equation} \label{times}
\tau _\ell = \frac {N^2}{ y_\ell ^2} \, \tau_{tw}
\end{equation}
gives the $l^{th}$ elastic relaxation time of the dsDNA region.
The longest relaxation time for $\lambda$-DNA
is $\tau_1 =3.5\,10^{-3}$ s.

If the number of unzipped base pairs $n(t)$
is small with respect to $N$, then (\ref{noft}) is valid.
For $\lambda$-DNA this condition 
happens to be true over the time range $0<t<\tau _1$ where $\tau _1$ is the
longest relaxation mode of the double helix [see (\ref{times})]. 
This means that the force will reach its peak at about $\tau_1$,
and then will not vary significantly for later times $\sim \tau_1$.
The dsDNA rotation dynamics reaches a stationary regime, and
at later times can be considered to be a rigid cylinder rotating 
at angular velocity $\omega = \theta _0 dn/dt$.   For times
beyond $\tau_1$, the twist-relaxation dynamics reduce to
just the differential equation for the number
of unzipped base pairs (\ref{dynan}).

To analyze the dynamics for $t < \tau_1$ we solve (\ref{integ}) 
iteratively.  We start from
the equilibrium zero speed solution $n_0(t) \propto t$. At step $i$, the
number of unzipped base pairs as a function of time, $n_i(t)$, 
is inserted in the r.h.s. of (\ref{integ}) and $n_{i+1} (t)$
is collected on the l.h.s.  The iteration is repeated until convergence
is obtained, which takes about 20 iterations.
We also solved the ordinary differential equation (\ref{dynan}) using numerical 
integration routines, and observed that the two curves match 
accurately for times $t \ge \tau _1$ as expected. 

\subsection{Results}

Fig.~9a shows force versus displacement including the twist 
relaxation dynamics for homogeneous sequence ($g_0 = 2.5 k_B T$).
The only difference with the curve obtained without 
the twist propagation (shown for comparison for the velocity of 
$20\ \mu$m/s, dashed curve) is the much smoother initial force increase.
Theory indicates a force increase spread over the first
$\simeq 2.5\ \mu$m of displacement; after this extension
the result converges to the one obtained without twist propagation.
While we do observe an initial `delay' of the force increase,
the range over which theory predicts this effect is
shorter than that observed experimentally (about $5\ \mu$m).

We have also studied the dynamics of the
opening  fork including both twist propagation and the
$\lambda$ sequence. In presence of a complex free energy landscape for $g_0$,
the iterative scheme exposed above does not converge easily to the solution
of (\ref{noft}). We have therefore consider smoother landscapes through
a preaveraging of the sequence over $\delta N$ bases, with $\delta N$ ranging
from 1000 down to 100. Such values permit to reach numerical convergence
and are sufficient to detect sequence--induced effects on the micron scale.   
The resulting force signal is shown in Fig.~9b for maximal velocity $2v=
20 \mu$m/s. The percentage of GC bases increases during the initial opening,
and this effect spreads the initial force increase over the initial 
$5 \mu$m of extension. This trend is in qualitative agreement with 
experimental findings, though the calculated force is lower than the 
experimental value by a few pN.

\section{Conclusion}

We have presented a theory of DNA unzipping dynamics,
for kilobase or longer dsDNAs. We have shown
that torsional drag built by the rotation of the double helix around its
axis is the dominant frictional contribution in the opening of
$\lambda$-DNA molecules.  Easily observable
nonequilibrium effects for $\lambda$-DNA
are expected for unzipping velocities 
in excess of 2 $\mu$m/s.  The results of our  theory
are in agreement with recent observations by Thomen et al\cite{heslot}
of a roughly 40\% increase in unzipping force for $\lambda$-DNA
unzipping at $2v = 20$ $\mu$m/s.  

We have neglected a few physical effects in the discussion above.  First, 
we have not explicitly included effects of transport of the dsDNA base-pairs 
{\it to} the fork.  This is particularly relevant to the experimental
setup of Refs. \cite{1,2,heslot} where the fork moves relative to one of
the ssDNA anchor points.  This may introduce an additional
translational contribution to the dsDNA drag (for a rod model,
again proportional to $N-n$).    However, an estimate made by
Thomen et al\cite{heslot} suggests that this force should be
small relative to the unzipping forces.
In fact, at the lower fork velocities $\approx 2$ $\mu$m/s where
nonequilibrium effects are observable, separate experiments and theory
shows that a $\lambda$-DNA coil should only be slightly stretched\cite{macro}.

A second factor that we have ignored is the possible effect of
dsDNA intrinsic bends.  Nelson has recently
argued that such bends should induce an orders-of-magnitude 
effective enhancement in the rotational drag coefficient \cite{23}.
In the present experiment, it appears that this effect is nearly
absent. Possibly, the roughly twofold enhancement of $r$
over its `bare' chemical value, needed to generate the observed
unzipping force enhancement, is due to permanent bends or other
structural inhomogeneities along the rotating double helix.

A third, and potentially interesting effect is that 
once appreciable torque is built up in the dsDNA, there is the
possibility that the upstream dsDNA may begin to writhe \cite{16}.
For a dsDNA under zero force, writhing (supercoiling) occurs
when $|\Gamma| > k_B T$ \cite{22}. Tension in excess of 
$k_B T/A \approx 0.1$ pN ($A=50$ nm is the dsDNA bending persistence
length) pushes the writhing threshold up to
$|\Gamma| > (4 k_B T A f)^{1/2}$.
Writhing (chiral coiling) of the dsDNA could increase the
effective $r$, even without the formation of plectonemic supercoils.
Formation of plectonemes is straightforward if the dsDNA region starts
as a random coil, since there will be near-crossings every few kb 
(every 5 to 6 persistence lengths) to act as plectoneme `anchors'.
Once plectonemes form, we expect a large enhancement in the
effective friction for dsDNA rotation, and a large increase in
unzipping force.

It would be very interesting to see results for
an experiment carried out in the geometry of Fig. 1. 
This might be done using two translated laser traps, which
would allow much less perturbation of the dsDNA coil during unzipping.

A natural question is raised by the absence of a noise term in 
eqn (\ref{dinfor}), preventing the system from  probing the whole 
free energy landscape at equilibrium\cite{Boc02} (see also
\cite{9bis}, Section VII). 
This approximation, which makes easier for the system to
be blocked in a local minimum (stick regime), is expected to be 
valid at large velocities {\em i.e.} when the landscape changes very 
fast, and times scales are too small to allow for barrier crossing
between stick and slip states. To further test the validity of 
eqn (\ref{dinfor}), we have calculated the number of open base pairs 
$n$ as a function of displacement $2x$ at extremely low velocity 
e.g. 10 nm/s for the $\lambda$ sequence. Results are in very good 
agreement with equilibrium predictions of Fig.~4; the only difference  
is that unstick jumps (Inset of Fig.~4) take sometimes place
$\le 5$ nm after their equilibrium counterparts.

Finally, we note that the intrinsic fork motion time $\tau_n$
might be larger than the $\approx 10^{-8}$/s
assumed in this paper using dimensional considerations.  
The activation barriers to opening of successive bases \cite{7,11} 
might make $\tau_n$ larger; a recent estimate based
on the analysis of a RNA opening experiment at constant force
\cite{5} gave $\tau _n \simeq 2\, 10^{-7}$s\cite{sfig}. 
However, there will be little consequence of a larger $\tau_n$
for the phenomena discussed in this paper, which occur on a
much longer time scale.

\section*{Acknowledgements}

This work was supported by NSF Grant DMR-9734178 and DMR-0203963,
by a Research Innovation Award from Research Corporation,
and by a Focused Giving Award from Johnson \& Johnson Corporate Research.

\begin{figure}
\begin{center}
\includegraphics[height=250pt,angle=0] {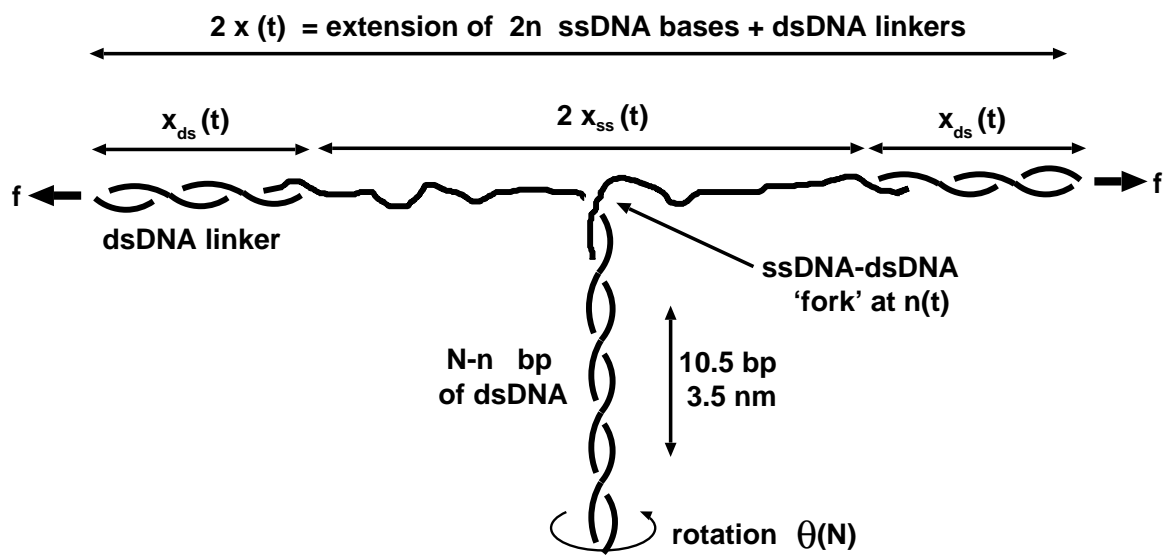} 
\end{center}
\caption{DNA unzipping experiment considered
in this paper.  Connections are made to dsDNA linkers attached
to the ssDNA ends, and are used to pull adjacent 3' and 5' ssDNA ends apart.
The ends of the dsDNA linkers move at equal and opposite velocities
of magnitude $v$.  As the dsDNA is converted to separated ssDNAs,
the helical turns of the dsDNA must be expelled, forcing the
remaining dsDNA to be rotated once for each 10.5 bases which
are unzipped. }
\end{figure}

\begin{figure}
\begin{center}
\includegraphics[height=300pt,angle=-90] {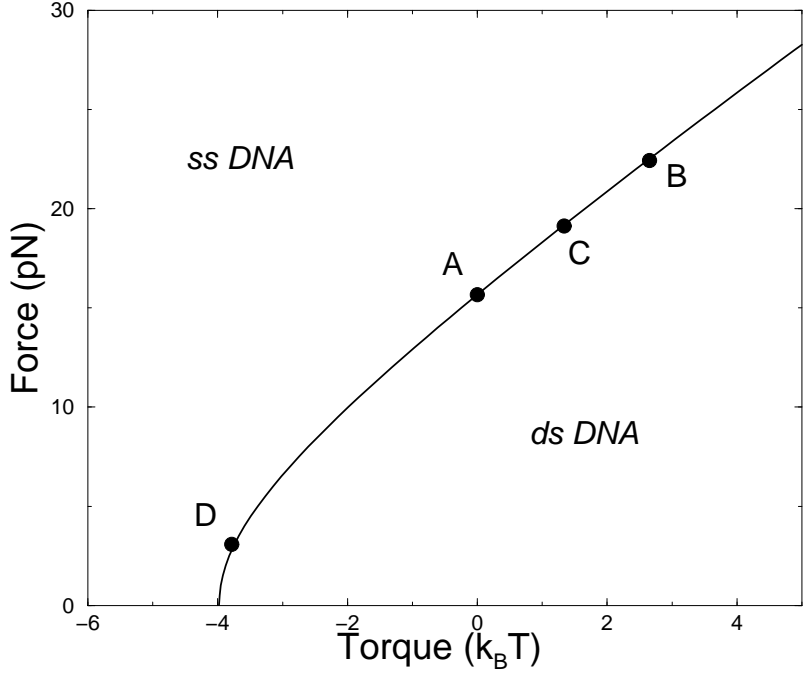} 
\end{center}
\caption{Phase diagram of a homogeneous DNA molecule 
under applied force and torque. The base pairing free energy is set
to $g_0=2.5 k_BT$. The double helix conformation is
thermodynamically preferred when the
applied force is smaller than some torque-dependent critical value,
e.g. $f_u=15.7$ pN at zero torque. Denaturation may be driven by
torque only. In the absence of applied force, DNA opens when an
underwinding torque larger (in modulus) than 4 $k_BT$ is applied.
Points A,B,C,D refer to the force vs. displacement curve of
Fig.~6.}
\end{figure}

\begin{figure}
\begin{center}
\includegraphics[height=300pt,angle=-90] {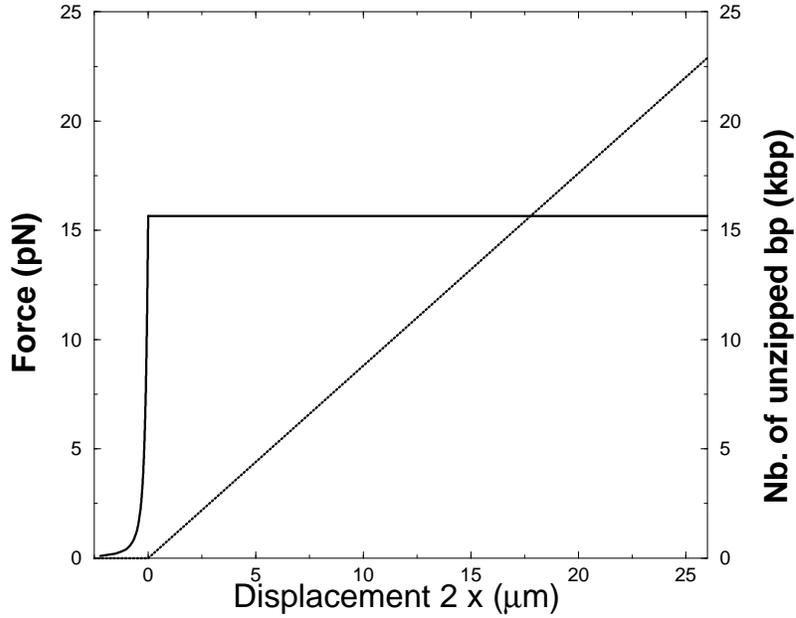} 
\end{center}
\caption{Force (pN,full curve) and number of unzipped base pairs 
(kbp, dotted curve) 
at equilibrium for a homogeneous DNA molecule, as a function
of the displacement $x$ after opening initiation. Prior to unzipping
($x<0$), the force vs. extension curve reflects the elastic behavior
of linkers.  As opening proceeds ($x>0$), force is constant
at $f_u=15.7$ pN, for base pairing energy $g_0=2.5 k_BT$. 
The number of unzipped base pairs increases linearly with $x$,
with a slope $\simeq 1$ bp/nm.}
\end{figure}

\begin{figure}
\begin{center}
\includegraphics[height=300pt,angle=-90] {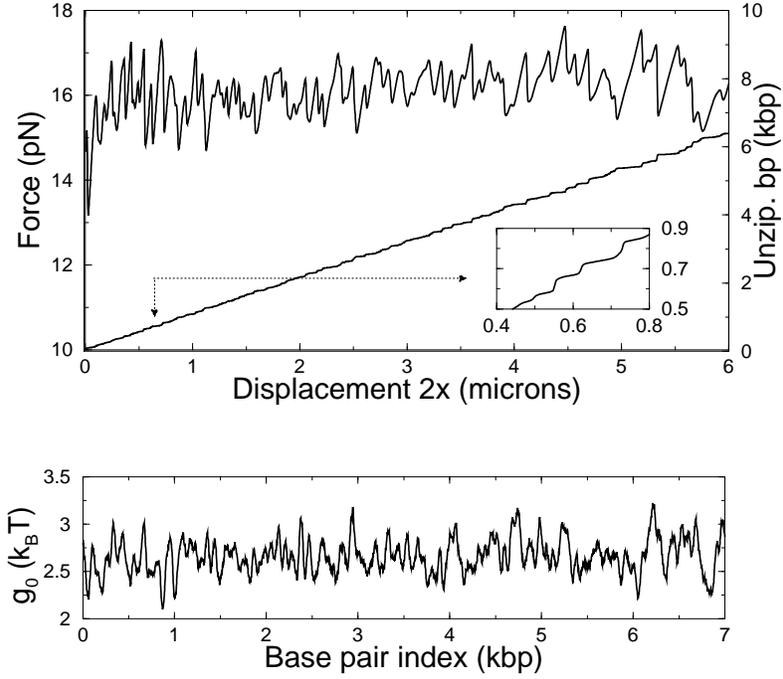} 
\end{center}
\caption{Force (pN, top curve) and number of unzipped base pairs 
(kbp, middle curve) at equilibrium for the $\lambda$--DNA molecule, 
as a function of the displacement $2x$ after opening initiation. 
Bottom curve is the base pairing free energy vs. index of base pair, 
from a Gaussian average over 20 bp.
The number of unzipped base pairs increases linearly with $x$, with
characteristic stick-slip steps (see inset).}
\end{figure}

\begin{figure}
\begin{center}
\includegraphics[height=250pt,angle=-90] {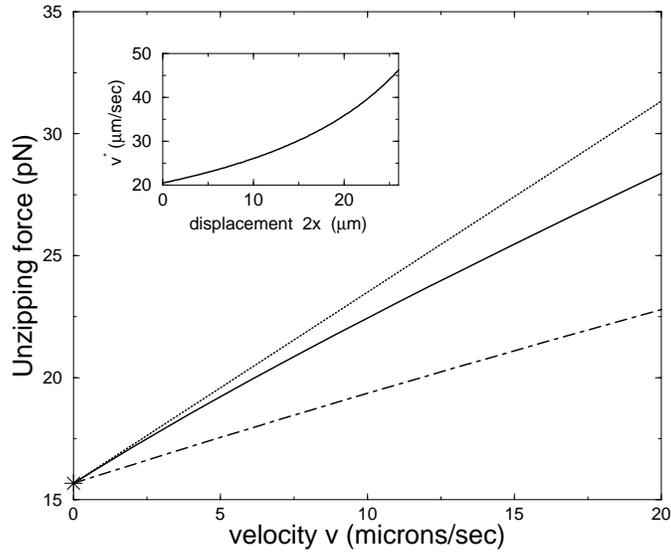}
\end{center}
\caption{Theoretical unzipping force as a function of
pulling velocity for homogeneous DNA.  
Star indicates equilibrium unzipping force.
Dashed curve shows the `steady-state'
approximate formula (\ref{newforce}), solid curve shows the peak force 
encountered during integration of (\ref{dynan}).  
The dashed and solid curves have
the same dependence at low velocity, tending to the static
unzipping force $f_u \approx 15.7$ pN at zero
velocity.  Inset shows characteristic velocity
$v*$ entering steady-state formula is plotted as a function of
displacement $2x$. For velocities above $v^*(0)=20$ $\mu$m/s,
the steady-state formula starts to be appreciably above
the peak force obtained by integration of (\ref{dynan}). 
The dot-dashed curve shows the force at the half-unzipped
point when about 25 kbp are unzipped.
}
\end{figure}

\begin{figure}
\begin{center}
\includegraphics[height=250pt,angle=-90] {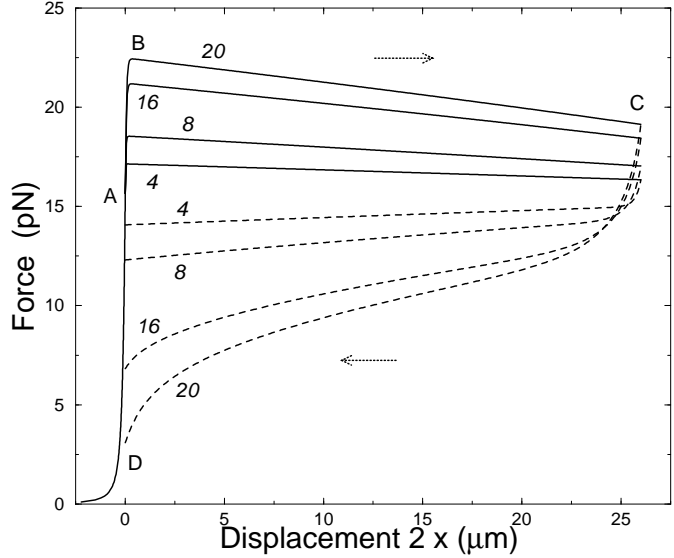}
\end{center}
\caption{Force vs. single-strand extension $x=v t$ obtained
from integration of (\ref{dynan}), for homogeneous DNA plus 7 kb linkers
as in Fig.~1.  Outgoing (pulling) curves for 
$2v=4$,  8, 16 and 20 $\mu$m/s are shown (solid curves, bottom to top).
For these rates, successively higher unzipping forces
are obtained.  In all cases an initial force increase associated
with pulling the linkers taut, is followed by a force peak,
and then a slow force reduction.  The slow and linear reduction
of force with extension is due to progressively less dsDNA
being left to provide rotational drag to oppose fork motion.
Force during retraction following the extensions are shown for
$2v=4$,  8, 16 and 20 $\mu$m/s (dashed curves).
For 2 and 4 $\mu$m/s, relatively small hysteresis loops occurs.  
However, at retraction at $\geq 8$ $\mu$m/s the hysteresis is
larger due to larger rotational drag,
the force drops near zero at the end of rezipping.}
\end{figure}

\begin{figure}
\begin{center}
\includegraphics[height=250pt,angle=-90] {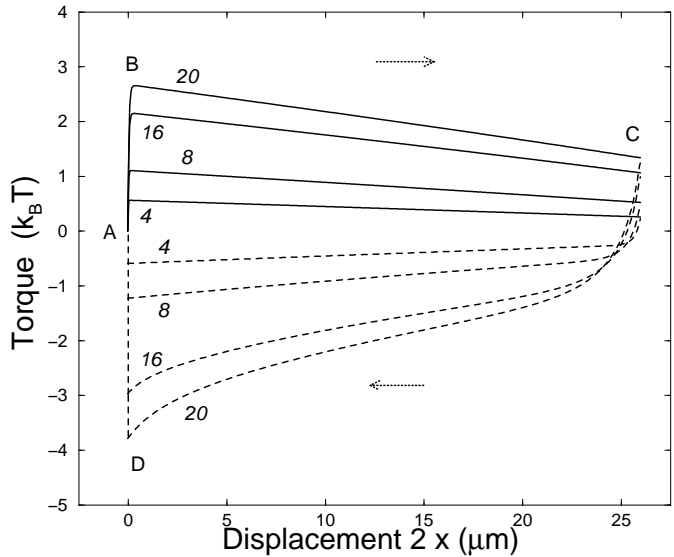}
\end{center}
\caption{Torque in the dsDNA region immediately adjacent
to the fork, vs single-strand extension $x=vt$, obtained
from integration of (\ref{dynan}), for the cases shown in Fig. 6.
Solid curves indicate extension, dashed retraction.
Progressively higher torques are obtained at successively
higher pulling velocities.  There is an appreciable peak in
the torque during the early stages of unzipping, followed
by a gradual torque decay as the remaining dsDNA provides
progressively less rotational drag.  This torque buildup
is due to the rapid fork motion forcing the dsDNA region
just upstream of the fork to be under overtwisting strain.
The torques obtained during retraction (dashed curves) show that
rewinding of the molecule generates 
left-handed elastic torque in the molecule; for retractions
of 8 and 16 $\mu$m/s this torque exceeds the `free' limit
of $g_0/\theta_0 \approx -4 k_B T$ discussed in the text.  }
\end{figure}

\begin{figure}
\begin{center}
\includegraphics[height=280pt,angle=-90] {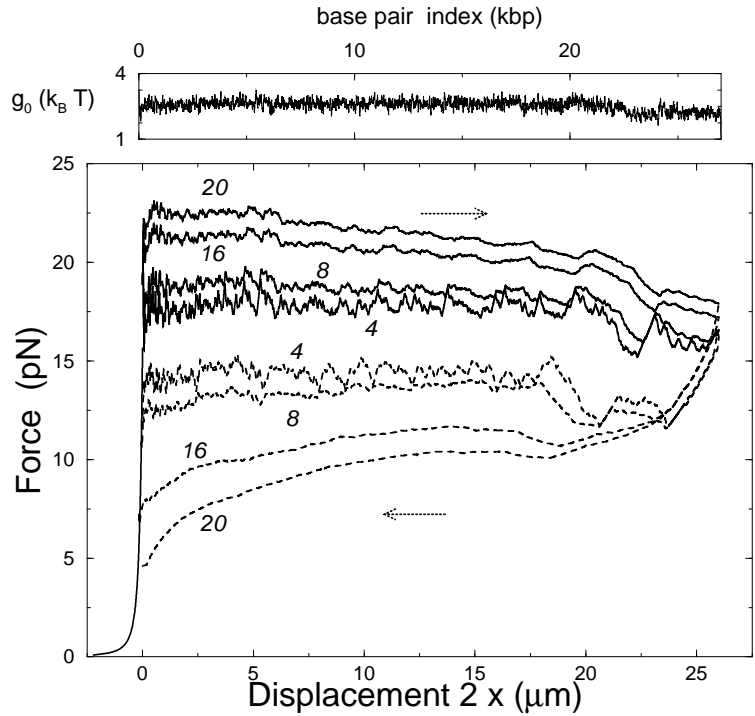}
\end{center}
\caption{Force vs. single-strand extension $x=v t$ for
$\lambda$--DNA from integration of (\ref{dynan}), for 
$2v=$ 4, 8, 16 and 20 $\mu$m/s (outgoing curves: full line,
retracting curves: dashed lines). Note that fluctuations of the force
due to sequence effects decrease with increasing velocities.}
\end{figure}

\begin{figure}
\begin{center}
\includegraphics[height=250pt,angle=-90] {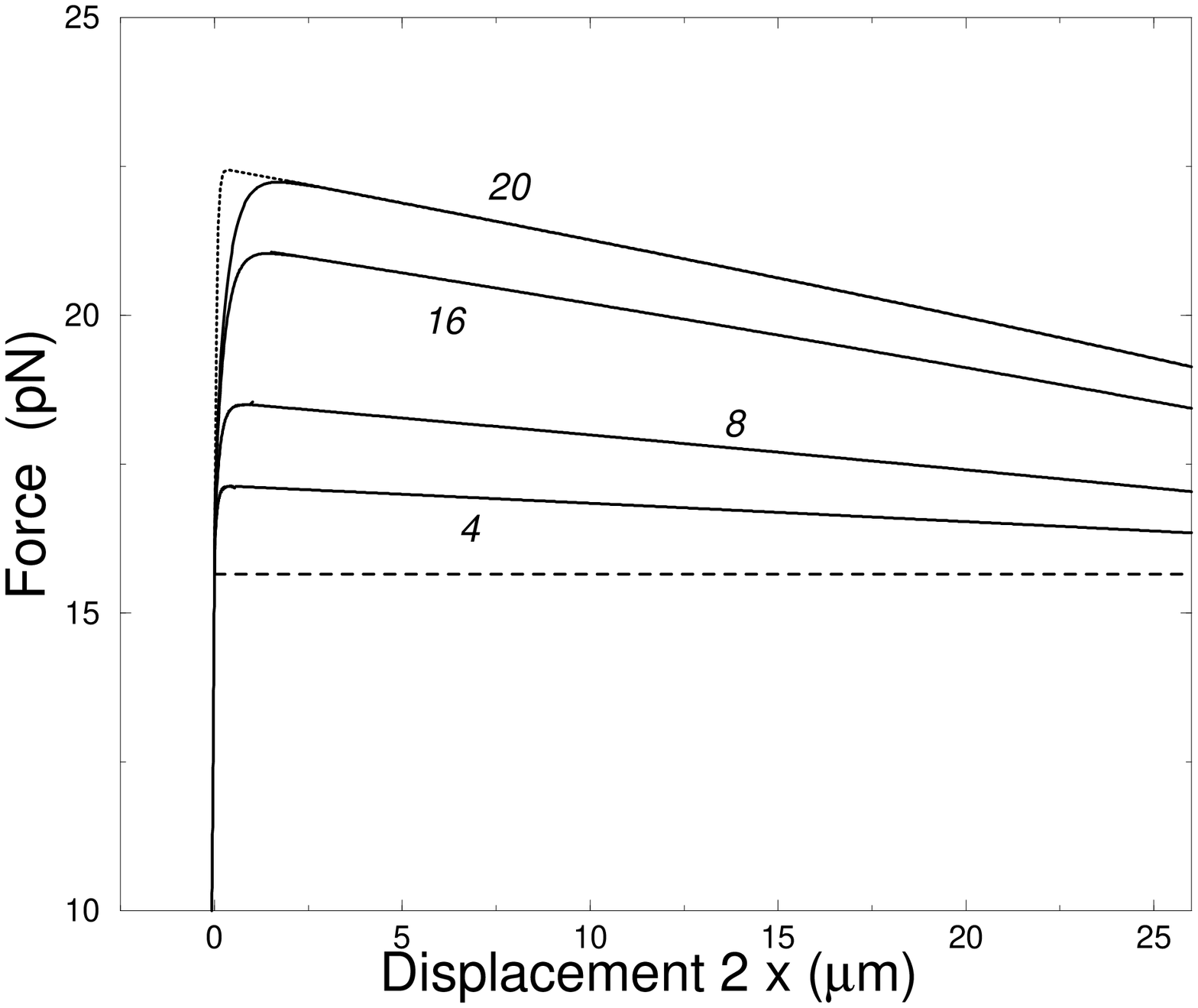}
\includegraphics[height=250pt,angle=-90] {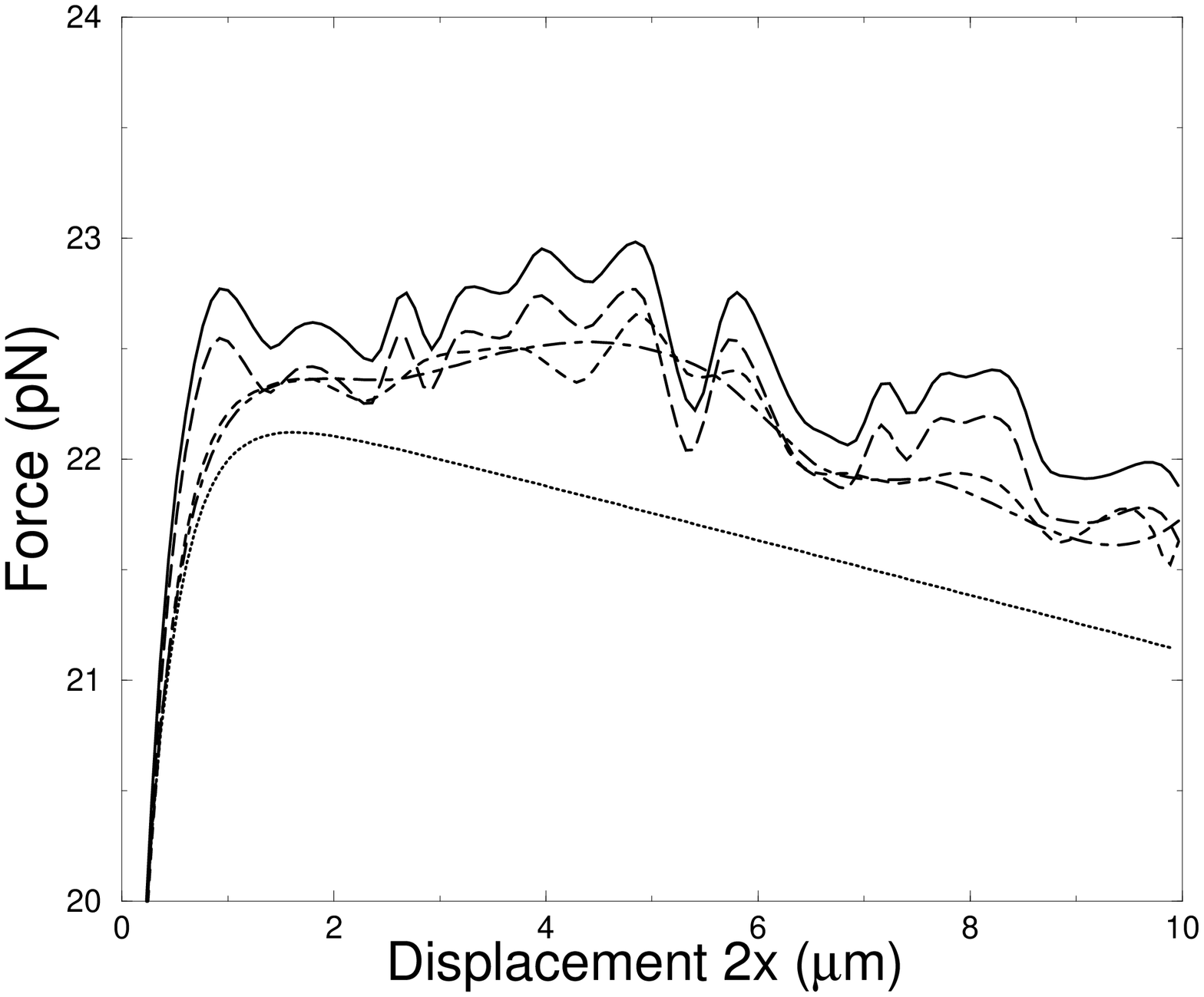}
\end{center}
\caption{Force vs. single-strand extension $x=v t$ with twist
propagation obtained from resolution of (\ref{noft}).
(a) Result for homogeneous 
DNA ($g_0=2.5 k_B T$) and with 7 kb linkers.
Outgoing (pulling) curves for 
$2v=2$, 4, 8, 16 and 20 $\mu$m/s are shown (solid curves, bottom to top).
For these rates, successively higher unzipping forces
are obtained.  The initial force increase associated
with pulling the linkers is much smoother than without twist
propagation (shown for $2v=20$ $\mu$m/s).
(b) Corresponding results for inhomogeneous sequence for $2v=20 \mu$m/s,
compared to the homogeneous case (dotted line). Curves correspond
to preaveraging of the sequence over 1000 (dot-dashed line), 500 (dashed),
250 (long-dashed) and 100 (solid) base pairs.}
\end{figure}


\begin{references}

\bibitem{1}
U. Bockelmann, B. Essevaz-Roulet, F. Heslot,
{\it Proc. Natl. Acad. Sci. USA} {\bf 94}, 11935-40 (1997).

\bibitem{2}
U. Bockelmann, B. Essevaz-Roulet, F. Heslot,
{\it Phys. Rev. Lett} {\bf 79},4489-442 (1997);
{\it Phys. Rev. E} {\bf 58}, 2386-2384 (1998).

\bibitem{3}
 M. Rief, H. Clausen-Schaumann, H.E. Gaub,
{\it Nat. Struct. Biol.} {\bf 6}, 346-9 (1999). 

\bibitem{4} 
C. Bustamante, S.B. Smith, J. Liphardt, D. Smith,
{\it Curr. Opin. Struct. Biol.} {\bf 10}, 279-85 (2000).

\bibitem{5} 
J. Liphardt, B. Oona, S.B. Smith, I. Tinoco, C. Bustamante,
{\it Science} {\bf 292}, 733-737 (2001).

\bibitem{6} 
J.L. Viovy, C. Heller, F. Caron, P. Cluzel, D. Chatenay,
{\it C.R. de l'Acad. des Sci.} {\bf 317}, 795-800 (1994).

\bibitem{7} 
R.E. Thompson, E.D. Siggia,
{\it Europhys. Lett.} {\bf 31} 335-40 (1995).

\bibitem{8} 
K.L. Sebastian,
{\it Phys. Rev. E} {\bf 62} 1128-32 (2000).

\bibitem{9}
D.K. Lubensky, D.R. Nelson,
{\it Phys. Rev. Lett.} {\bf 85}, 1572-5 (2000).

\bibitem{9bis}
D.K. Lubensky, D.R. Nelson,
{\it Phys. Rev. E} {\bf 65}, 031917 (2002).

\bibitem{10} 
S.M. Bhattacharjee,
{\it J. Phys. A} {\bf 33}, L423 (2000).

\bibitem{11} 
S. Cocco, R. Monasson, J.F. Marko,
{\it Proc. Natl. Acad. Sci. USA} {\bf 98}, 8608-13 (2001);
S. Cocco, J.F. Marko, R. Monasson,
{\em Phys. Rev. E} {\bf 65}, 041907 (2002).

\bibitem{12} 
U. Gerland, R. Bundschuh, T. Hwa,
{\it Biophys. J.} {\bf 81}, 1324-32 (2001).

\bibitem{13} 
T. Strick, J.-F. Allemand, D. Bensimon, R. Lavery, V. Croquette,
{\it Physica A} {\bf 263}, 392-404 (1999).

\bibitem{14} 
S. Cocco, R. Monasson,
{\it Phys. Rev. Lett.} {\bf 83} 5178-81 (1999).

\bibitem{heslot}
P. Thomen, U. Bockelmann and F. Heslot,
{\it Phys. Rev. Lett.} {\bf 88}, 248102 (2002).

\bibitem{15} 
C. Levinthal, H.R. Crane,
{\it Proc. Natl. Acad. Sci. USA} {\bf 42}, 436 (1956).

\bibitem{16} 
J.F. Marko, 
{\it Phys. Rev. E} {\bf 57}, 2134-49 (1998).

\bibitem{17}
A. Sarkar, J.F. Marko,
{\it Phys. Rev. E} {\bf 64}, 1909 (2001).

\bibitem{Zuk00}
M. Zuker, {\em Curr. Opin. Struct. Biol} {\bf 10}, 303 (2000).

\bibitem{San98}
J. Santa Lucia, {\it Proc. Natl. Acad. Sci. USA} {\bf 95}, 1460-65 (1998).

\bibitem{Wang}
M. Wang, H. Yin, R. Landick, J. Gelles, S.M. Block,
{\it Biophys. J.} {\bf 72}, 1335-1346 (1997).

\bibitem{18} 
S.B. Smith, Y. Cui, C. Bustamante,
{\it Science} {\bf 271}, 795 (1996).

\bibitem{20}
K.J. Breslauer, R. Frank, H. Blocker, L.A. Marky, 
{\it Proc. Natl. Acad. Sci. USA} {\bf 83}, 3746-3750 (1986).

\bibitem{Boc02}
U. Bockelmann, P. Thomen, B. Essevaz-Roulet, V. Viasnoff, F. Heslot.
{\it Biophys. J.} {\bf 82}, 1537-1553 (2002).

\bibitem{21}
L.D. Landau, E.M. Lifshitz,
{\it Theory of Elasticity} (Pergamon, New York, 1986), Sec. 16.

\bibitem{22}
J.F. Marko, E.D. Siggia,
{\it Phys. Rev. E} {\bf 52}, 2912-2938 (1995).

\bibitem{macro}
J.F. Marko, E.D. Siggia,
{\it Macromol.} {\bf 28}, 8759 (1995).

\bibitem{23}
P. Nelson,
{\it Proc. Natl. Acad. Sci. USA} {\bf 96}, 14342-47 (1999).

\bibitem{sfig}
S. Cocco, J.F. Marko, R. Monasson, Slow nucleic acid unzipping
kinetics from sequence-defined barriers, {\em preprint} (2002).

\end{references}
\end{document}